\documentclass[prl,twocolumn,showpacs,aps]{revtex4-1}

\usepackage{amsmath}
\usepackage{amssymb}
\usepackage{txfonts}
\usepackage{graphicx}
\usepackage{epsfig}
\usepackage{enumerate}

\usepackage{ifpdf}
\ifpdf
\usepackage{epstopdf}
\fi

\usepackage[unicode]{hyperref}
\hypersetup{
	pdftitle={MultiSolutions},
	pdfauthor={Nian, Qian},
	pdfsubject={MultiSolutions},
	colorlinks=true,
	linkcolor=blue,
	citecolor=blue,
	filecolor=black,
	urlcolor=blue
}

\def\url#1{}

\usepackage{bm}% bold math
\usepackage{color}
\newcommand{\be}{\begin{equation}}

\newcommand{\ee}{\end{equation}}

\newcommand{\bea}{\begin{eqnarray}}

\newcommand{\eea}{\end{eqnarray}}

\begin{document}

\title{Exact Multi-Instanton Solutions to Selfdual Yang-Mills Equation on Curved Spaces}
\author{Jun Nian$^{1}$}
\email{nian@umich.edu}
\author{Yachao Qian$^{2}$}
\email{yachao.qian@alumni.stonybrook.edu}
\affiliation{$^1$ Leinweber Center for Theoretical Physics, University of Michigan, Ann Arbor, MI 48109, USA}
\affiliation{$^2$ Department of Physics and Astronomy, Stony Brook University, Stony Brook, NY 11794-3800, USA}

\begin{abstract}

We find exact multi-instanton solutions to the selfdual Yang-Mills equation on a large class of curved spaces with $SO(3)$ isometry, generalizing the results previously found on $\mathbb{R}^4$. The solutions are featured with explicit multi-centered expressions and topological properties. As examples, we demonstrate the approach on several different curved spaces, including the Einstein static universe and $\mathbb{R} \times$ dS$_3^E$, and show that the exact multi-instanton solutions exist on these curved backgrounds.

\end{abstract}

\maketitle

{\flushleft\bf 1.~Introduction}

To precisely study a quantum theory, the first step is to find its vacuum structure. The well-known example is the Yang-Mills instanton solution \cite{instanton}, which plays an important role in the research of non-perturbative aspects of quantum field theories, for instance, in the 4d $\mathcal{N}=2$ Seiberg-Witten prepotential \cite{SW} from instanton counting \cite{Nekrasov}. Moreover, instantons have featured prominently in recent attempts at finding the low-energy effective theory and addressing some long-standing problems of Yang-Mills theory \cite{Shuryak, YM-1}.

In recent years, the technique of supersymmetric localization has been applied to many supersymmetric gauge theories on curved spaces to obtain some exact quantities such as the partition function. The key idea is to integrate quantum fluctuations at quadratic order around the saddle-point configurations. In some cases, these saddle-point configurations are solutions to the (anti-)selfdual Yang-Mills equation, hence are (anti-)instanton or (anti-)vortex solutions near the poles of the sphere \cite{Pestun, S2-1, S2-2, Ellipsoid, 3dHiggs, N=1}. Therefore, the first step of supersymmetric localization also relies on finding exact solutions to the selfdual Yang-Mills equation on curved spaces.

After the first nontrivial solution constructed in \cite{instanton}, there have been more nontrivial solutions to Yang-Mills equation found in the literature (see e.g. \cite{Actor:1979in}). Among the various nontrivial solutions to Yang-Mills equation, the exact multi-instanton solution on $\mathbb{R}^4$ found by Witten \cite{multi-instanton} is particularly interesting, because it has an explicit expression for higher topological numbers and obeys the cylindrical symmetry instead of the spherical symmetry obeyed by the standard instanton solution \cite{instanton}.

Besides the solutions on flat spaces, there are also nontrivial solutions found on curved spaces \cite{BoutalebJoutei:1979va, BoutalebJoutei:1979ec, BoutalebJoutei:1979rw, BoutalebJoutei:1979iz, BoutalebJoutei:1979iy, Comtet:1983wt, Comtet:1979kq, Verbin-1, Verbin-2, Tekin:2002mt, Nielsen, Lechtenfeld-1, Lechtenfeld-2, Lechtenfeld-3}. These spherically symmetric solutions on curves spaces have been systematically analyzed in the same manner as \cite{Nian:2019vbm}. A natural question is whether the exact multi-instanton solution in \cite{multi-instanton} can be generalized to curved spaces. For some special curved spaces, one can indeed find the exact multi-instanton solution case by case, for instance, $\mathbb{R} \times \mathbb{H}^3$ discussed in \cite{Harland:2007cq} and the Euclidean AdS$_4$ black hole discussed in \cite{Sarioglu:2009du}. In this paper, we resolve this problem through a systematic way by constructing the exact multi-instanton solutions on a large class of curved spaces and studying their topological properties. As we will show, by this systematic approach, new exact multi-instanton solutions on the Einstein universe and $\mathbb{R} \times$ dS$_3^E$ can be easily obtained, both of which have not been considered in the literature before. Also, we can prove that such solutions have integer topological charges.

{\flushleft\bf 2.~Selfdual Yang-Mills Equation on Curved Space}

We consider a large class of curved Euclidean spaces $\mathbb{R} \times \mathcal{M}_3$ given by the following metric:
\be\label{eq:metric}
  ds^2 = dt^2 + h(r)\, dr^2 + r^2\, d\Omega_{S^2}^2\, ,
\ee
where $r$ is the radial coordinate in the spatial part, and $t$ denotes the Euclidean time, and the 3d manifold $\mathcal{M}_3$ has at least an $SO(3)$ isometry. For example, the conformally flat Euclidean space $\mathbb{R} \times S^3$ is also in the class \eqref{eq:metric}. Hence, one can find exact multi-instanton solutions on $\mathbb{R} \times S^3$ by mapping the solutions on $\mathbb{R}^4$ \cite{multi-instanton} conformally to $\mathbb{R} \times S^3$, which is not possible for general spaces in the class \eqref{eq:metric}.

The metric \eqref{eq:metric} can be written into some equivalent expressions:
\begin{align}
\begin{split}\label{eq:metricNew}
  ds^2 & = dt^2 + dr^2 + r^2\, d\Omega_{S^2}^2 + (h - 1) dr^2 \\
  {} & = dx_\mu\, dx^\mu + \frac{h - 1}{r^2} (\vec{x}\cdot d\vec{x})^2\, ,
\end{split}
\end{align}
i.e. the non-vanishing components of the metric $g_{\mu\nu}$ are
\be\label{eq:g_mu nu}
  g_{00} = 1\, ,\quad g_{ij} = \delta_{ij} + \frac{h-1}{r^2} x_i x_j\, ,
\ee
where $x_\mu$ and $x^\mu$ ($\mu = 1,\cdots, 4$) are pseodu-Cartesian coordinates. We can construct them in the following way: rewriting the original metric \eqref{eq:metric} into the expression \eqref{eq:metricNew}, we embed the two-sphere $S^2$ into a 3d flat space $\mathbb{R}^3$ with standard Cartesian coordinates $x_i$ ($i = 1, 2, 3$), and $r$ becomes the radial coordinate in $\mathbb{R}^3$. Together with the Euclidean time direction $t$ in the original metric \eqref{eq:metric}, we obtain a 4d flat space  $\mathbb{R}^4$ with Cartesian coordinates $x^\mu$ ($\mu \in \{1, \cdots, 4\}$) with $x^4 = t$, whose indices can be raised or lowered using the Kronecker delta. For convenience, we also define $x^0 \equiv t = x^4$. The advantage of using the pseudo-Cartesian coordinates is that for a curved space the metric \eqref{eq:metric} can be decomposed into a flat part and a curved part, as shown in the second line of \eqref{eq:metricNew}. From the components \eqref{eq:g_mu nu}, we can compute
\be
  \textrm{det}\, g = h\, .
\ee

%and the non-vanishing components of the inverse metric:
%\be
%  g^{00} = 1\, ,\quad g^{ij} = \delta^{ij} + \frac{1 - h}{r^2 h} x^i x^j\, .
%\ee

For the gauge group $SU(2)$, we adopt the following Ansatz for the components of the gauge field:
\begin{align}
\begin{split}\label{eq:Ansatz}
  A_{j, a} & = \frac{\varphi_2 + 1}{r^2} \epsilon_{j a k} x_k + \frac{\varphi_1}{r^3} (\delta_{j a} r^2 - x_j x_a) + \frac{A_1}{r^2} x_j x_a\, ,\\
  A_{0, a} & = \frac{A_0 x_a}{r}\, .
\end{split}
\end{align}
The field strength is defined as
\be
  F_{\mu\nu, a} \equiv \partial_\mu A_{\nu, a} - \partial_\nu A_{\mu, a} - \epsilon_{abc} A_{\mu, b} A_{\nu, c}\, .
\ee
Consequently, the components of the field strength can be obtained as follows:
\begin{align}
  F_{0 i, a} & = (\partial_0 \varphi_2 - A_0 \varphi_1) \frac{\epsilon_{i a k} x_k}{r^2} + (\partial_0 \varphi_1 + A_0 \varphi_2) \frac{\delta_{ia} r^2 - x_i x_a}{r^3} \nonumber\\
  {} & \quad + r^2 (\partial_0 A_1 - \partial_r A_0) \frac{x_i x_a}{r^4}\, ,\label{eq:F_0i}\\
  \frac{1}{2} \epsilon_{ijk} F_{jk, a} & = - (\partial_r \varphi_1 + A_1 \varphi_2) \frac{\epsilon_{iak} x_k}{r^2} + (\partial_r \varphi_2 - A_1 \varphi_1) \frac{\delta_{ia} r^2 - x_i x_a}{r^3} \nonumber\\
  {} & \quad + (1 - \varphi_1^2 - \varphi_2^2) \frac{x_i x_a}{r^4}\, ,\label{eq:F_ij}
\end{align}
where the derivative $\partial_r$ is given by $\partial/\partial x_i = \left(\partial r / \partial x_i\right)\, \partial_r = \left(x_i/r\right)\, \partial_r$. All the indices in field strength except $r$ can be raised using the Kronecker delta, while for the coordinate $r$ we should use the factor $h$ in the original metric \eqref{eq:metric} to raise or lower it. Similar to the flat space $\mathbb{R}^4$, now for the curved space \eqref{eq:metric} the Yang-Mills action can still be effectively reduced to a 2d Abelian Higgs model:
\begin{align}
  {} & S = \frac{1}{4} \int dt\, \int d^3 x\, \sqrt{h}\, F_{\mu\nu, a} F_{\mu\nu, a} \nonumber\\
  = & \,\, 8 \pi \int dt \int dr \sqrt{h}\, \Bigg[\frac{1}{2} (D_\alpha \varphi_i)^2 + \frac{r^2}{8} F_{\alpha\beta}^2 + \frac{1}{4 r^2} \left(1 - \varphi_1^2 - \varphi_2^2 \right) \Bigg]\, ,\label{eq:Action}
\end{align}
where $\alpha, \beta \in \{1, 2\}$ and we define $\partial_1 \equiv \partial_r$. The physical configurations should have finite actions, which can be used as a condition on the factor $h$ and consequently as a criterium for the existence of physical solutions on a curved space given by the metric \eqref{eq:metric}.

The selfdual Yang-Mills equation on the curved space given by the metric \eqref{eq:metric} is
\be
  F_{0i, a} = \frac{1}{2 \sqrt{g}} \epsilon_{ijk} F_{jk, a} = \frac{1}{2 \sqrt{h}} \epsilon_{ijk} F_{jk, a}\, .
\ee
Applying the expressions \eqref{eq:F_0i} and \eqref{eq:F_ij} of the field strength, we can rewrite the selfdual Yang-Mills equation into a system of partial differential equations
\begin{align}
\begin{split}\label{eq:YM components}
  \partial_0 \varphi_2 - A_0 \varphi_1 & = - \frac{1}{\sqrt{h}} \partial_r \varphi_1 - \frac{1}{\sqrt{h}} A_1 \varphi_2\, ,\\
  \partial_0 \varphi_1 + A_0 \varphi_2 & = \frac{1}{\sqrt{h}} \partial_r \varphi_2 - \frac{1}{\sqrt{h}} A_1 \varphi_1\, ,\\
  r^2 \left(\frac{1}{\sqrt{h}} \partial_0 A_1 - \frac{1}{\sqrt{h}} \partial_r A_0 \right) & = 1 - \varphi_1^2 - \varphi_2^2\, .
\end{split}
\end{align}
Defining
\begin{align}
\begin{split}\label{eq:A1tilde, rtilde}
  \widetilde{A}_0 \equiv A_0\, ,\quad \widetilde{r}^2 \widetilde{A}_1 \equiv & \frac{r^2}{\sqrt{h}} A_1\, ,\quad \widetilde{r}^2 \partial_{\widetilde{r}} \equiv \frac{r^2}{\sqrt{h}} \partial_r\, ,\\
  \widetilde{\varphi}_0 = \varphi_0\, , & \quad \widetilde{\varphi}_1 = \varphi_1\, ,
\end{split}
\end{align}
we can rearrange the equations \eqref{eq:YM components} into the following form:
\begin{align}
\begin{split}\label{eq:YM components New}
  \partial_{\widetilde{r}}\, \widetilde{\varphi}_1 - \widetilde{A}_0\, \widetilde{\varphi}_1 & = - (\partial_0\, \widetilde{\varphi}_2 + \widetilde{A}_1\, \widetilde{\varphi}_2)\, ,\\
  \partial_0\, \widetilde{\varphi}_1 + \widetilde{A}_1\, \widetilde{\varphi}_1 & = \partial_{\widetilde{r}}\, \widetilde{\varphi}_2 - \widetilde{A}_0\, \widetilde{\varphi}_2\, ,\\
  \widetilde{r}^2 \left(\partial_0 \widetilde{A}_1 - \partial_{\widetilde{r}} \widetilde{A}_0 \right) & = 1 - \widetilde{\varphi}_1^2 - \widetilde{\varphi}_2^2\, .
\end{split}
\end{align}
Using the definitions \eqref{eq:A1tilde, rtilde}, these equations look formally the same as the equations on $\mathbb{R}^4$ considered in \cite{multi-instanton}.

{\flushleft\bf 3.~Exact Multi-Instanton Solution}

As we have seen, after some redefinitions \eqref{eq:A1tilde, rtilde}, the equations obtained from the tensor decomposition of selfdual Yang-Mills equation have expressions similar to the $\mathbb{R}^4$ case discussed in \cite{multi-instanton}. Hence, we can apply the same method to find the exact multi-instanton solutions for the curved space \eqref{eq:metric}.

First, we choose the gauge fixing condition $\partial^\mu \widetilde{A}_\mu = 0$, which implies that there exists a function $\psi$ such that
\be\label{eq:redefineA}
  \widetilde{A}_\alpha = \epsilon_{\alpha\beta} \partial_\beta \psi\, ,\quad (\alpha,\, \beta \in \{0, 1\})
\ee
where we also define $\partial_1 \equiv \partial_{\widetilde{r}}$. Moreover, instead of $\widetilde{\varphi}_i$ we define two new functions $\chi_i$ as
\be\label{eq:define chi}
  \chi_i \equiv e^{- \psi}\, \widetilde{\varphi}_i\, ,\quad (i \in \{0, 1\})\, .
\ee
In terms of $\chi_i$, the first two equations of \eqref{eq:YM components New} can be written as
\be
  \partial_0 \chi_1 = \partial_1 \chi_2\, ,\quad \partial_1 \chi_1 = - \partial_0 \chi_2\, ,
\ee
or more compactly as
\be
  \partial_{\bar{z}} f = 0\, ,
\ee
where
\be\label{eq:define z and f}
  z \equiv \widetilde{r} + i t\, ,\quad f \equiv \chi_1 + i \chi_2\, .
\ee

Using the expression \eqref{eq:redefineA}, we can also rewrite the third equation of \eqref{eq:YM components New} as
\be\label{eq:Third Eq of YM components}
  - \widetilde{r}^2 \nabla^2 \psi = 1 - \widetilde{\varphi}_1^2 - \widetilde{\varphi}_2^2 = 1 - e^{2 \psi}\, f^* f\, ,
\ee
which is invariant under the following transformations with an arbitrary holomorphic function $h(z)$ without singularity:
\begin{align}
  f \rightarrow f\, h\, ,\quad \psi \rightarrow \psi - \frac{1}{2} \textrm{ln}\, (h^* h)\, .
\end{align}
However, since the equation \eqref{eq:Third Eq of YM components} can also be written as
\be
  - \nabla^2 \psi = \frac{1}{\widetilde{r}^2} - \frac{1}{\widetilde{r}^2} f^*\, f\, e^{2 \psi}\, ,
\ee
which has a singularity at $\widetilde{r}=0$ from the first term on the right-hand side, we can also choose a non-holomorphic function $h(z)$ to cancel the term $1/ \widetilde{r}^2$, for instance, $h(z)$ with $|h| = \widetilde{r}^{\textrm{ln}\, \widetilde{r}}$. Hence, we only need to consider the remaining equation without singularities, i.e.,
\be\label{eq:New Third Eq of YM components}
  \nabla^2 \psi = \frac{1}{\widetilde{r}^2} f^*\, f\, e^{2 \psi}\, .
\ee
To solve this new equation, let us adopt the Ansatz
\be\label{eq:psi Ansatz}
  \psi = \textrm{ln}\, \widetilde{r} - \frac{1}{2} \textrm{ln} (f^*\, f) + \rho\, .
\ee
Plugging it into the equation \eqref{eq:New Third Eq of YM components}, we obtain the Liouville equation for $\rho$:
\be\label{eq:Regular Third Eq of YM components}
  \nabla^2 \rho = e^{2 \rho}\, .
\ee

The solution to \eqref{eq:Regular Third Eq of YM components} is formally given by
\be
  \rho = - \textrm{ln} \left[\frac{1}{2} (1 - g^* g) \right] + \frac{1}{2}\, \textrm{ln} \bigg|\frac{dg}{dz} \bigg|^2\, ,
\ee
where $g(z)$ is an analytic function. We can choose $g(z)$ such that $dg / dz = f$. Therefore, the solution to $\psi$ is
\be\label{eq:psi and f}
  \psi = - \textrm{ln} \left[\frac{1 - g^* g}{2 \widetilde{r}} \right]\, ,\quad \textrm{with}\quad \frac{d g}{d z} = f\, .
\ee
In order that $\psi$ is nonsingular, $g$ should satisfy
\begin{align}
\begin{split}
  |g| = 1\, ,& \quad \textrm{at}\quad \widetilde{r} = 0\, ;\quad |g| < 1\, ,\quad \textrm{for}\quad \widetilde{r} > 0\, ;\\
  g \textrm{ finite}\, ,& \quad \textrm{at}\quad \widetilde{r} = \infty\, .
\end{split}
\end{align}
The most general solution of $g$ obeying these boundary conditions is
\be\label{eq:g}
  g = \prod_{i=1}^k \frac{a_i - z}{a_i^* + z}\, ,\quad (a_i: \textrm{ complex constants})
\ee
which characterizes the exact multi-instanton solutions on the curved spaces \eqref{eq:metric}.

To summarize, we have found the new coordinates $(t, \widetilde{r})$ and the new fields $(\widetilde{\varphi}_i,\, \widetilde{A}_i)$, such that in these new coordinates the multi-instanton solutions can be obtained from the ones on $\mathbb{R}^4$ \cite{multi-instanton}. A solution is first characterized by the function $g$ \eqref{eq:g}, then $\psi$ and $f$ can be computed using \eqref{eq:psi and f}. Based on the definition \eqref{eq:define z and f}, we can read off $\chi_i$ from the real and the imaginary parts of $f$. Finally, $\widetilde{A}_i$ and $\widetilde{\varphi}_i$ can be obtained from \eqref{eq:redefineA} and \eqref{eq:define chi} respectively, and all of them are functions of $(t,\, \widetilde{r})$.

The solutions on different backgrounds in the class \eqref{eq:metric} all look the same in the coordinates $(t,\, \widetilde{r})$, and the difference enters when writing the solutions in terms of the original components defined in \eqref{eq:Ansatz}. In order to express the solutions as \eqref{eq:Ansatz}, we should solve the last equation of \eqref{eq:A1tilde, rtilde} to obtain $\widetilde{r} (r)$, and subsequently relate $\widetilde{A}_1$ with $A_1$ using the second equation of \eqref{eq:A1tilde, rtilde}, i.e.,
\begin{align}
\begin{split}
  \varphi_i (r) & = \widetilde{\varphi}_i \Big( \widetilde{r} (r) \Big)\, ,\quad A_0 (r) = \widetilde{A}_0 \Big(\widetilde{r} (r) \Big)\, ,\\
  {} & A_1 (r) = \frac{\sqrt{h}}{r^2}\, \widetilde{r}^2 (r)\, \widetilde{A}_1 \Big(\widetilde{r} (r) \Big)\, .
\end{split}
\end{align}
We see that the multi-instanton solutions on curved spaces return to the solution on $\mathbb{R}^4$ in the flat-space limit, i.e., $h = 1$ and $\widetilde{r} = r$.

{\flushleft\bf 4.~Topological Property}

To discuss the topological property of the multi-instanton solution on the curved spaces \eqref{eq:metric}, we introduce the topological term:
\be\label{eq:Top term}
  n = \frac{1}{8 \pi^2} \int d^4 x \sqrt{h}\, F_{\mu\nu} \left(*F_{\mu\nu}\right)\, ,
\ee
where $*F_{\mu\nu}$ is the dual field strength. Using the explicit expressions of the components \eqref{eq:F_0i} \eqref{eq:F_ij}, we can express the topological term \eqref{eq:Top term} as
\be\label{eq:2d top term}
  n = \frac{1}{2 \pi} \int dt\, dr \, \Big[\epsilon_{\alpha\beta} \epsilon_{ij} D_\alpha \varphi_i D_\beta \varphi_j + \frac{1}{2} \epsilon_{\alpha\beta} F_{\alpha\beta} (1 - \varphi_1^2 - \varphi_2^2) \Big]\, .
\ee
Both terms in the intergrand above can be written as total derivatives, hence only boundary configurations contribute. If the configurations satisfy $D_\mu \varphi_i = 0$ at the boundary (see Sec.~5), the only non-vanishing contribution in \eqref{eq:2d top term} is
\be
  n = \frac{1}{4 \pi} \int dt\, dr \, \epsilon_{\alpha\beta} F_{\alpha\beta} = \frac{1}{4 \pi} \int dt\, d\widetilde{r} \, \epsilon_{\alpha\beta} \widetilde{F}_{\alpha\beta}\, ,
\ee
where we have used \eqref{eq:A1tilde, rtilde} and \eqref{eq:redefineA}. This expression can be further brought into the form
\be\label{eq:top term final}
  n = \frac{1}{2 \pi i} \oint ds\, \frac{d}{ds} \textrm{ln} \widetilde{\varphi}\, ,
\ee
where we define $\widetilde{\varphi} \equiv \widetilde{\varphi}_1 + i\, \widetilde{\varphi}_2 = f\, e^\psi$ based on \eqref{eq:define chi}. Since $\psi$ is continuous and nonsingular, \eqref{eq:top term final} essentially computes the number of zeroes of $f$, i.e., $n = k - 1$, which can be interpretted as the instanton number.

{\flushleft\bf 5.~Examples}

Let us discuss some examples in more detail.
\begin{enumerate}[(1)]

\item Einstein static universe:

The metric of the 4d Euclidean Einstein static universe is
\be
  ds^2 = dt^2 + d\theta^2 + \textrm{sin}^2 \theta\, d\Omega_2^2\, .
\ee
Defining $r \equiv \textrm{sin}\, \theta$, we can rewrite the metric as
\be\label{eq:EuclideanEinstein}
  ds^2 = dt^2 + \frac{1}{1 - r^2} dr^2 + r^2 \Omega_2^2\, ,\quad r \in [0,\, 1]\, ,
\ee
i.e. $h(r) = 1 / (1 - r^2)$ in this case. Solving \eqref{eq:A1tilde, rtilde}, we obtain for the 4d Euclidean Einstein static universe:
\be
  \widetilde{r} = \frac{r}{\sqrt{1 - r^2}}\, ,\quad A_1 (r) = \frac{1}{(1 - r^2)^{3/2}}\, \widetilde{A}_1 \left(\frac{r}{\sqrt{1 - r^2}} \right)\, ,
\ee
and correspondingly for the other fields
\begin{align}
\begin{split}
  \varphi_1 (r) = \widetilde{\varphi}_1 \left(\frac{r}{\sqrt{1 - r^2}}\right)\, , & \quad \varphi_2 (r) = \widetilde{\varphi}_2 \left(\frac{r}{\sqrt{1 - r^2}}\right)\, ,\\
  A_0 (r) = & \widetilde{A}_0 \left(\frac{r}{\sqrt{1 - r^2}}\right)\, .
\end{split}
\end{align}
As an example, the numerical result of $F^2$ for a 4-instanton solution is shown in Fig.~\ref{fig:Graph1}. We also plot $\textrm{log} (F^2)$ in Fig.~\ref{fig:logGraph1}, which clearly shows that $F^2$ remains finite and has no divergences at the instanton locations.

In this case, both $D_\alpha \varphi_i$ and $\sqrt{h}\, (D_\alpha \varphi_i)^2$ vanish at the boundary $r \to 1$, so do $\sqrt{h}\, r^2 F^2$ and $\sqrt{h}\, (1 - \varphi_1^2 - \varphi_2^2) / r^2$. The action \eqref{eq:Action} is finite for this case. Applying the analysis in Sec.~4, we see that the topological number is indeed equal to the multi-instanton number as expected.

   \begin{figure}[!htb]
      \begin{center}
        \includegraphics[width=0.42\textwidth]{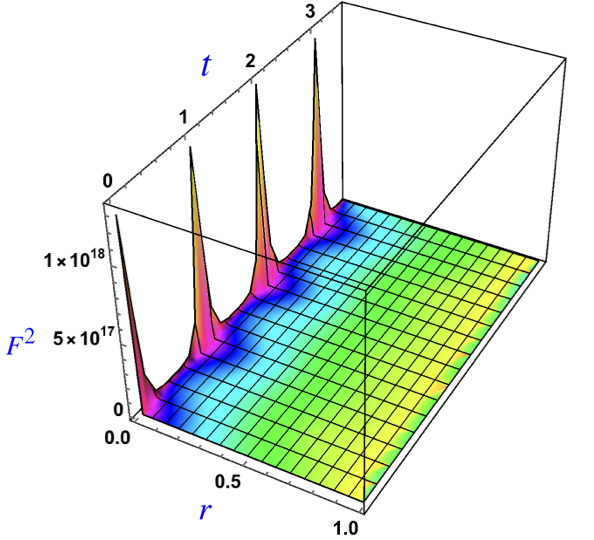}
      \caption{$F^2$ for a 4-instanton solution in Einstein static universe}
      \label{fig:Graph1}
      \end{center}
    \end{figure}

   \begin{figure}[!htb]
      \begin{center}
        \includegraphics[width=0.4\textwidth]{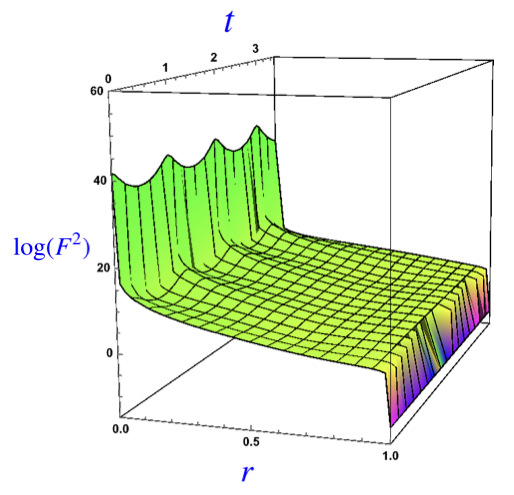}
      \caption{$\textrm{log} (F^2)$ for a 4-instanton solution in Einstein static universe}
      \label{fig:logGraph1}
      \end{center}
    \end{figure}

\item $\mathbb{R} \times$ dS$_3^E$:

The Euclidean dS$_3$ space has the following metric:
\be\label{eq:dS3}
  ds_{3d}^2 = b^2 \left(d\eta_E^2 + \textrm{cos}^2 \eta_E\, d\Omega_2^2 \right)\, .
\ee

Defining $r \equiv b\, \textrm{cos}\, \eta_E$, we can rewrite the metric \eqref{eq:dS3} into the form:
\be
  ds_{3d}^2 = \frac{dr^2}{1 - \frac{r^2}{b^2}} + r^2\, d\Omega_2^2\, ,\quad r \in [0,\, b]\, .
\ee
The 4d space $\mathbb{R} \times$ dS$_3^E$ is then given by the metric:
\be
  ds_{4d}^2 = dt^2 + \frac{dr^2}{1 - \frac{r^2}{b^2}} + r^2\, d\Omega_2^2\, ,\quad r \in [0,\, b]\, ,
\ee
which is a special case in the class \eqref{eq:metric} with the factor $h(r) = (1 - r^2 / b^2)^{-1}$. Solving \eqref{eq:A1tilde, rtilde}, we obtain for $\mathbb{R} \times$ dS$_3^E$:
\be
  \widetilde{r} = \frac{b\, r}{\sqrt{b^2 - r^2}}\, ,\,\, A_1 (r) = \frac{b^3}{(b^2 - r^2)^{3/2}}\, \widetilde{A}_1 \left(\frac{b\, r}{\sqrt{b^2 - r^2}} \right)\, ,
\ee
and correspondingly for the other fields
\begin{align}
\begin{split}
  \varphi_1 (r) = \widetilde{\varphi}_1 \left(\frac{r}{\sqrt{1 - r^2}}\right)\, , & \quad \varphi_2 (r) = \widetilde{\varphi}_2 \left(\frac{r}{\sqrt{1 - r^2}}\right)\, ,\\
  A_0 (r) = & \widetilde{A}_0 \left(\frac{r}{\sqrt{1 - r^2}}\right)\, .
\end{split}
\end{align}
As an example, the numerical result of $F^2$ for a 4-instanton solution with $b = 0.1$ is shown in Fig.~\ref{fig:Graph3}. Again, we plot $\textrm{log} (F^2)$ in Fig.~\ref{fig:logGraph3} to show that $F^2$ remains finite and has no divergences at the instanton locations.

In this case, both $D_\alpha \varphi_i$ and $\sqrt{h}\, (D_\alpha \varphi_i)^2$ vanish at the boundary $r \to b$, so do $\sqrt{h}\, r^2 F^2$ and $\sqrt{h}\, (1 - \varphi_1^2 - \varphi_2^2) / r^2$. The action \eqref{eq:Action} is finite for this case. Applying the analysis in Sec.~4, we see that the topological number is indeed equal to the multi-instanton number as expected.

   \begin{figure}[!htb]
      \begin{center}
        \includegraphics[width=0.42\textwidth]{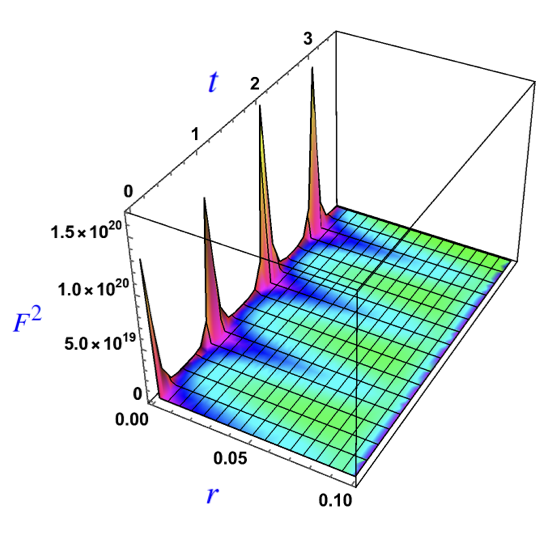}
      \caption{$F^2$ for a 4-instanton solution in $\mathbb{R} \times$ dS$_3^E$ with $b = 0.1$}
      \label{fig:Graph3}
      \end{center}
    \end{figure}

   \begin{figure}[!htb]
      \begin{center}
        \includegraphics[width=0.4\textwidth]{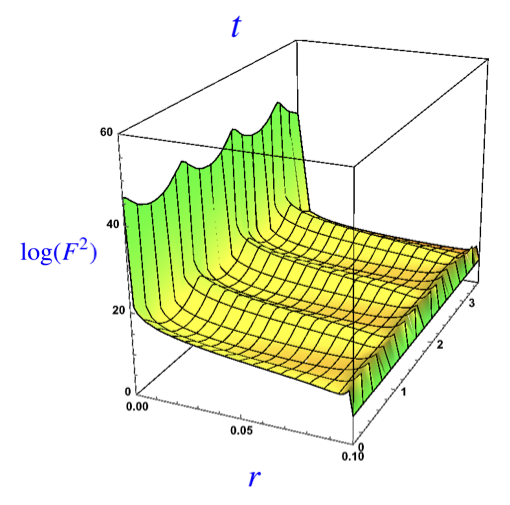}
      \caption{$\textrm{log} (F^2)$ for a 4-instanton solution in $\mathbb{R} \times$ dS$_3^E$ with $b = 0.1$}
      \label{fig:logGraph3}
      \end{center}
    \end{figure}

\end{enumerate}

\par
{\flushleft\bf 6.~Conclusions}

In this paper we have constructed the exact multi-instanton solution to Yang-Mills equation on a large class of curved spaces given by the metric \eqref{eq:metric}. By finding an appropriate new coordinate system and field redefinitions, we see that the Yang-Mills equation on the curved spaces \eqref{eq:metric} becomes the one on $\mathbb{R}^4$ preserving the cylindrical symmetry. Hence, the exact multi-instanton solutions on this class of curved spaces can be constructed in the same way as the flat space, which was discussed in \cite{multi-instanton}. We demonstrated the finite actions of these solutions, and discussed their topological properties.

As we have seen, the class of curved spaces includes some special examples (e.g. the Einstein static universe, $\mathbb{R} \times$ dS$_3^E$, etc.), which are important in study of cosmology. Hence, the exact multi-instanton solutions on these curved spaces may lead to observable physical effects. The exact multi-instanton solutions also appear as vacuum solutions to supersymmetric gauge theories on some curved spaces, which are crucial for the problems of the instanton counting or the vortex counting of supersymmetric gauge theories \cite{Pestun, S2-1, S2-2, Ellipsoid, 3dHiggs, N=1}.

Some generalizations to higher dimensions are also possible, which will extend the results in \cite{Fairlie:1984mp, Fubini:1985jm, Harvey:1990eg} and lead to exact octonionic multi-instanton solutions and exact multi-centered string solitons.

\par
{\bf Acknowledgments.---}
J.N.'s work was supported in part by the U.S. Department of Energy under grant DE-SC0007859 and by a Van Loo Postdoctoral Fellowship, and he would like to thank the New High Energy Theory Center at Rutgers University and Arizona State University for hospitality during the final stages of this work.

J.N. and Y.Q. contribute equally to this work.

\bibliographystyle{utphys}
\bibliography{MultiCurved}

\end{document}